\documentclass[sigplan,screen]{acmart}
\usepackage{threeparttable}
\usepackage{multirow}
\usepackage[linesnumbered,ruled,vlined]{algorithm2e}
\usepackage{microtype}
\usepackage{listings}
\usepackage{subcaption}
\usepackage{geometry}
\lstdefinestyle{mystyle}{
    backgroundcolor=\color{yellow!10},   
    commentstyle=\color{green!70!black},  
    keywordstyle=\color{blue},            
    numberstyle=\tiny\color{gray},        
    stringstyle=\color{red},              
    basicstyle=\ttfamily\footnotesize,    
    breakatwhitespace=false,               
    breaklines=true,                       
    numbers=left,                         
    numbersep=5pt,                        
    tabsize=2,                            
    frame=lines,                          
    framerule=1pt,                        
    rulecolor=\color{black},              
    aboveskip=2pt,                        
    belowskip=2pt                         
}

\def\BibTeX{{\rm B\kern-.05em{\sc i\kern-.025em b}\kern-.08em
    T\kern-.1667em\lower.7ex\hbox{E}\kern-.125emX}}
\usepackage{balance}
\AtBeginDocument{%
  \providecommand\BibTeX{{%
    Bib\TeX}}}

\setcopyright{acmlicensed}

\acmDOI{10.1145/3735452.3735526}

\acmYear{2025}

\copyrightyear{2025}

\acmISBN{979-8-4007-1921-9/25/06}

\acmConference[LCTES '25]{Proceedings of the 26th ACM SIGPLAN/SIGBED International Conference on Languages, Compilers, and Tools for Embedded Systems}{June 16--17, 2025}{Seoul, Republic of Korea}

\acmBooktitle{Proceedings of the 26th ACM SIGPLAN/SIGBED International Conference on Languages, Compilers, and Tools for Embedded Systems (LCTES '25), June 16--17, 2025, Seoul, Republic of Korea}

\received{2025-03-06}

\received[accepted]{2025-04-21}

\begin{document}
\title[Zoozve: A Strip-Mining-Free RISC-V Vector Extension with Arbitrary Register Grouping ...]{Zoozve: A Strip-Mining-Free RISC-V Vector Extension with Arbitrary Register Grouping Compilation Support (WIP)}

\titlenote{This work was supported in part by the National Natural Science Foundation of China (NSFC) under Grants 62271300 and 12141107, in part by the Shanghai Science and Technology Plan Project under Grants 24DP1501100 and 24DP1500600.}

\author{Siyi Xu}
\orcid{0009-0008-0764-0621}
\authornotemark[2]
\author{Limin Jiang}
\orcid{0009-0008-7034-5780}
\authornote{Both authors contributed equally to this research.}
\affiliation{%
  \institution{Shanghai University, Shanghai, China}
  \country{ }
}
\email{{xusiyi,jianglimin}@shu.edu.cn}

\author{Yintao Liu}
\orcid{0009-0002-3538-4676}
\affiliation{%
  \institution{Shanghai University}
  \city{Shanghai}
  \country{China}
}
\email{berialest@shu.edu.cn}

\author{Yihao Shen}
\orcid{0009-0006-2813-3448}
\affiliation{%
  \institution{Shanghai University}
  \city{Shanghai}
  \country{China}
}
\email{shenyihao@shu.edu.cn}

\author{Yi Shi}
\orcid{0000-0002-3240-7900}
\affiliation{%
  \institution{Shanghai University}
  \city{Shanghai}
  \country{China}
}
\email{yishi1996@shu.edu.cn}

\author{Shan Cao}
\orcid{0000-0003-3713-8671}
\affiliation{%
  \institution{Shanghai University}
  \city{Shanghai}
  \country{China}
}
\email{cshan@shu.edu.cn}

\author{Zhiyuan Jiang}
\orcid{0000-0002-8522-5721}
\affiliation{%
  \institution{Shanghai University}
  \city{Shanghai}
  \country{China}
}
\email{jiangzhiyuan@shu.edu.cn}



\begin{abstract}
Vector processing is crucial for boosting processor performance and efficiency, particularly with data-parallel tasks. The RISC-V "V" Vector Extension (RVV) enhances algorithm efficiency by supporting vector registers of dynamic sizes and their grouping. Nevertheless, for very long vectors, the static number of RVV vector registers and its power-of-two grouping can lead to performance restrictions. To counteract this limitation, this work introduces Zoozve, a RISC-V vector instruction extension that eliminates the need for strip-mining. Zoozve allows for flexible vector register length and count configurations to boost data computation parallelism. With a data-adaptive register allocation approach, Zoozve permits any register groupings and accurately aligns vector lengths, cutting down register overhead and alleviating performance declines from strip-mining. Additionally, the paper details Zoozve's compiler and hardware implementations using LLVM and SystemVerilog. Initial results indicate Zoozve yields a minimum 10.10$\times$ reduction in dynamic instruction count for fast Fourier transform (FFT), with a mere 5.2\% increase in overall silicon area.

\end{abstract}




\maketitle

\vspace{-0.4cm}
\section{Introduction}


Extensive computational needs have spurred advancements in vector instruction set architectures (ISAs). To better serve an array of computational requirements, modern vector extension technologies have gradually moved towards variable-length registers, which allow vector lengths to be dynamically adjusted to suit varying workloads. New vector ISAs, such as the Scalable Vector Extension (SVE) \cite{stephens2017arm} and the RISC-V ``V'' Vector Extension (RVV), have been introduced. These designs allow vectors to be resized dynamically according to the demands of particular computational tasks, thus providing additional flexibility. The objectives in designing SVE and RVV include optimizing performance and enhancing resource optimization by adapting to workload variations \cite{pohl2019performance}. In contrast to traditional fixed-length vector registers, these modern vector extensions not only eliminate compatibility conflicts between hardware and software but also markedly enhance computational efficiency. Numerous companies and universities have developed diverse vector extensions supporting various RVV release versions, catering to different areas such as high-performance computing (HPC) \cite{perotti2024ara2, minervini2023vitruvius, perotti2022new, cavalcante2019ara, schuiki2020stream}, neural networks \cite{andri2020extending}, \cite{louis2019towards}, the internet of things (IoT), and edge computing \cite{chen2020xuantie}, \cite{gautschi2017near}. 

The RVV extension faces intrinsic obstacles in domain-specific computations, such as wireless communications and artificial intelligence, which often involve ultra-long vectors. When vector length multipliers (LMULs) are larger, they limit the number of registers available for allocation, which increases register pressure and results in more register spilling. Conversely, smaller LMULs require frequent strip-mining, negatively affecting performance \cite{lai2022efficient}. Moreover, developers must possess an in-depth comprehension of RVV's varied functionalities and operations and need to optimize register usage and memory access to devise efficient vectorized code, adding to the complexity of programming. Therefore, enhancing performance requires meticulous kernel optimization specific to applications, balancing LMUL settings and register availability, and understanding architectural details.

This paper presents \textit{Zoozve}, a RISC-V vector extension that removes the need for strip-mining, improving performance for ultra-long vector operations through enhanced instruction formats and expanded register access.
This work includes several contributions, outlined as follows:

{\setlength{\parindent}{0pt}
\textbf{Strip-Mining-Free Vector Extension:} To address the fixed register count and power-of-two register group issues found in RVV, a flexible RISC-V vector instruction extension without strip-mining is proposed. 

\textbf{Arbitrary Register Grouping Strategy:} A register allocation strategy that adapts dynamically to the circumstances is introduced, intelligently modulating the distribution of registers in accordance with real-time vector lengths and the current register availability. 

\textbf{Compilation Support:} Intrinsic splitting and assembly coalescing passes have been developed, together with a comprehensive compilation mechanism for Zoozve using LLVM, allowing effortless conversion from high-level code to efficient machine instructions. 
}


\begin{figure}[!t]
  \centering
  \includegraphics[width=0.9\linewidth]{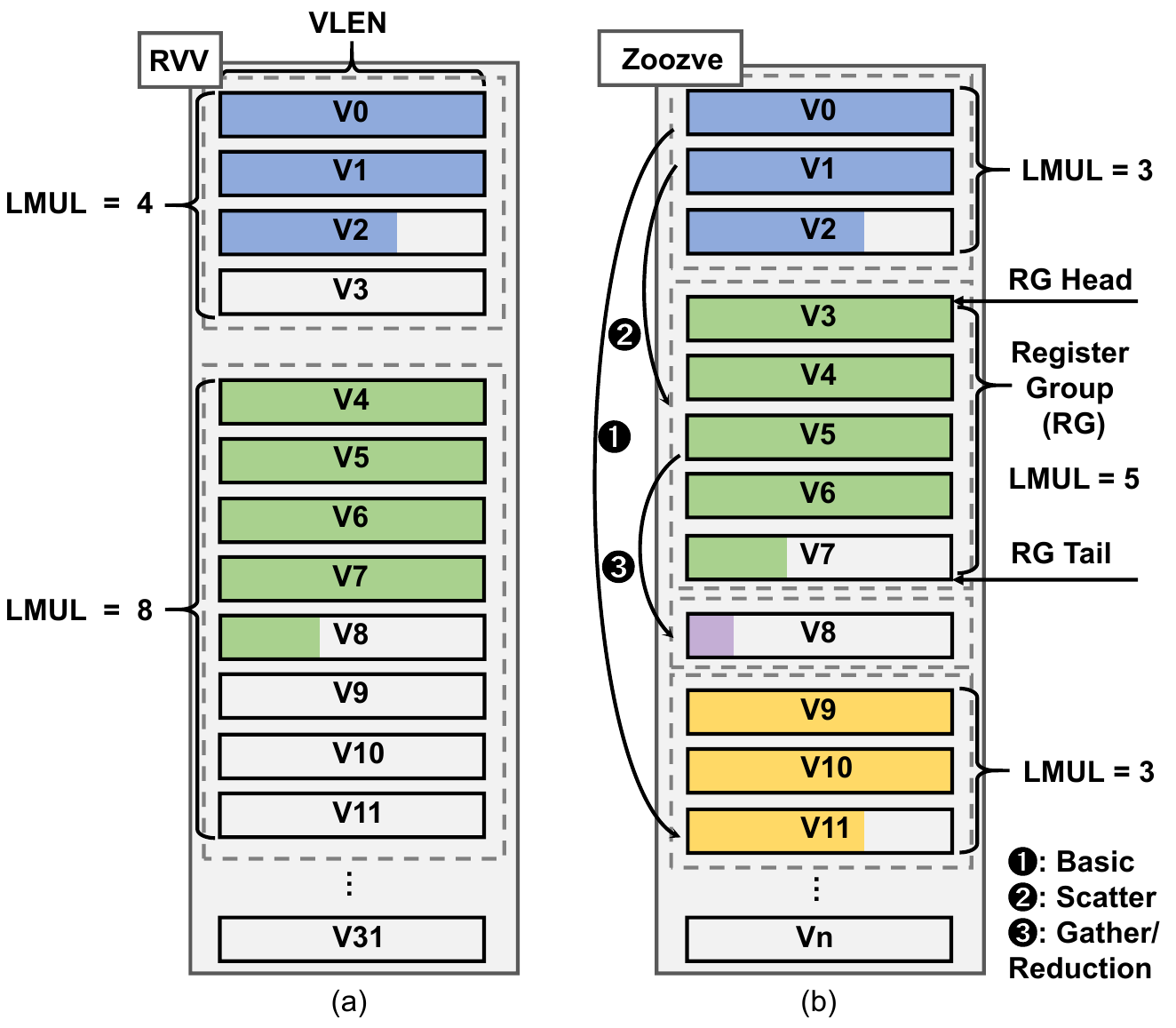}
  \vspace{-0.2cm}
  \caption{(a) Valid LMUL values for RVV include 4 and 8 across different vector lengths. (b) Zoozve supports arbitrary register grouping values for asymmetric instructions.}
  \vspace{-1cm}
  \label{fig:RegGroup2}
\end{figure}

\section{Background and Motivation}

{\setlength{\parindent}{0pt}
\textbf{Vector Strip-Mining: }
A fundamental aspect of vector ISAs is strip-mining, which enables vector processors to handle data volumes exceeding the capacity of available registers. This method partitions sizable vectors into smaller \textit{strips}, each handled separately within a loop, which can be implemented in either hardware or software \cite{hennessy2011computer}. For example, Advanced Vector eXtensions (AVX) \cite{lomont2011introduction} implements strip-mining through software without having specific hardware control registers. In contrast, SVE uses the \texttt{whilelt} predicative instruction for loop termination management. TSUBASA \cite{necaurora} and RVV enhance strip-mining by integrating hardware registers that dictate appropriate vector lengths for each strip. Crafting an ISA that minimizes conditional overhead while optimizing data-level parallelism demands careful consideration.

\textbf{Vector Register Grouping: }
Data structures for RVV and Zoozve are depicted in Fig.~\ref{fig:RegGroup2}. The power-of-two register grouping (RG) approach in RVV has limitations in two areas: (i) For relatively short vector lengths (VL), even though an LMUL can fit all vector elements into a single RG, a VL falling between $(2^{n_{lmul}-1}+1) \cdot \frac{VLEN}{VEW}$ and $(2^{n_{lmul}}-1) \cdot \frac{VLEN}{VEW}$ may cause under-utilization of vector registers. Here, $n_{lmul}$, $VLEN$, and $VEW$ correspond to the logarithm of LMUL, the bit width of an individual vector register, and the vector element bit width, respectively (illustrated in Fig.~\ref{fig:RegGroup2}(a)); (ii) In cases of longer VLs, RVV often faces challenges managing tail data during strip-mining, where small-sized tail data can occupy RGs in a higher LMUL setup, leading to diminished performance.
}

\begin{figure*}[!t] 
    \centering
    \includegraphics[width=\linewidth]{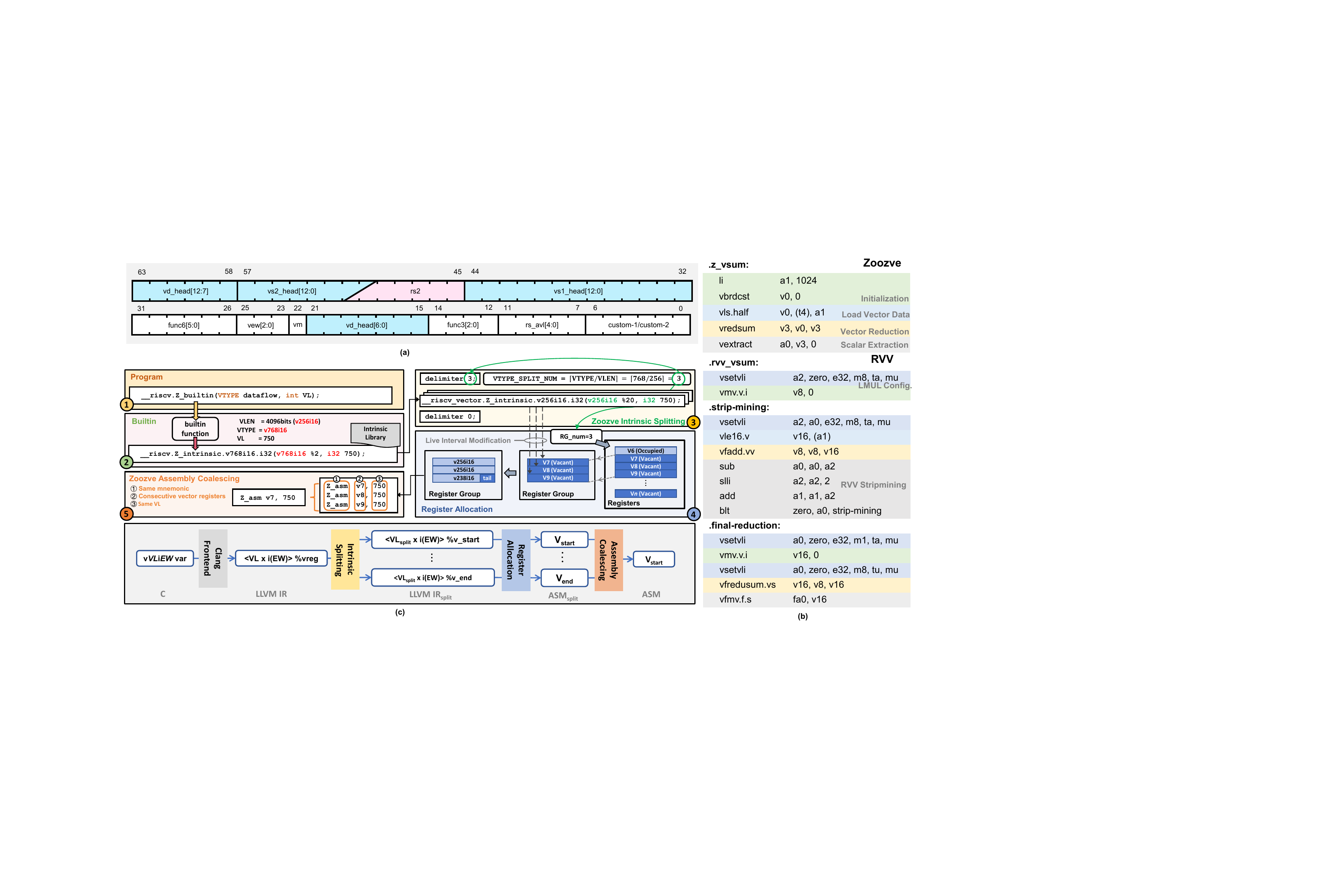}
    \vspace{-0.7cm}
    \caption{(a) Zoozve vector arithmetic and logical instructions format. (b) Comparison of reduction add assembly in Zoozve and RVV. (c) Compilation workflow for Zoozve in LLVM.}
    \label{fig:bigpic}
    \vspace{-0.3cm}
 \end{figure*}  

\section{ISA Extensions and Methodology}
\subsection{Zoozve Extension Instructions}
To address the aforementioned vector register grouping drawbacks, Zoozve is specifically designed for performant vector processing with the following key principles:
\textbf{(a)} The quantity and dimensions of vector registers can be arranged with flexibility, not limited to set values.
\textbf{(b)} Having an adequate quantity of vector registers enables more data to be kept in close proximity to the execution units.

Zoozve vector instructions are broadly categorized into three types: vector load/store instructions, vector arithmetic and logical instructions, and vector control instructions using RISC-V custom opcode regions (custom-0/1/2). Fig.\ref{fig:bigpic}(a) presents the typical vector arithmetic and logical instruction format. To support the efficient access of a large number of vector register groups, the \textbf{v\_head} field (e.g., \textbf{vd\_head}, \textbf{vs2\_head}, and \textbf{vs1\_head}) is used. This register field allows for the access of up to $2^{13}$ vector registers. It can be further expanded by using \texttt{vsetcsr} instruction to write extra bits into control and status registers (CSRs).  The field \textbf{v\_head} stores the starting addresses of the vector registers involved in the operation. The scalar register \textbf{rs\_avl} holds the target vector length. With the starting addresses provided by \textbf{v\_head} and the vector length specified by \textbf{rs\_avl}, it becomes possible to efficiently access large register groups. The \textbf{rs2} field is utilized to hold the scalar operand, enabling efficient vector-scalar computations. Fig.~\ref{fig:bigpic}(b) compares Zoozve and RVV in a reduction add kernel, highlighting Zoozve’s reduced instruction count due to the removal of strip-mining.

Asymmetric instructions are specifically designed for vector data processing, exemplified by scatter and gather instructions. Zoozve allows different VLs between source and destination vectors by using register-level gather or register scatter instructions, shown on the right of Fig.~\ref{fig:RegGroup2}. For lengthening vectors, a scatter instruction can be used: \texttt{vd[vs2[i]] $\leftarrow$ vs1[i]}. Conversely, to shorten a vector, a gather instruction can be applied: \texttt{vd[i] $\leftarrow$ vs1[vs2[i]]}. Unlike the \texttt{vrgather} instruction in RVV, which maintains the same VL for both source and target registers, Zoozve allows the target vector's length to match the length of the input indices, rather than the input vector's length, which minimizes register waste. These asymmetric instructions maximize register efficiency when there is a substantial VL difference between source and destination registers.

Leveraging the extensive encoding space described above, an arbitrary, data-adaptive register grouping allocation strategy is proposed to eliminate strip-mining at both the software and compiler level. 
Specifically, Zoozve allows for a variable number of vector registers ($\text{V}_{\text{0}}$ to $\text{V}_{\text{n}}$), which can be flexibly configured based on specific application requirements. 
The RGs in Zoozve can be dynamically adjusted based on the types of instruction values. The determination of RG depends on two factors: the starting register number, $RG_{head}$, allocated by register allocation (RA) algorithms of the compiler, and the vector data type, defined as $RG_{type}=L \cdot VEW$, where $L$ is the programming vector length. Data types are declared within the high-level programming language. At compile time, RGs are allocated by calculating the range from $RG_{head}$ to $RG_{tail} = RG_{head} + RG_{type}/VLEN$.

\subsection{Compilation Workflow}
To support Zoozve, we design a compilation workflow shown in Fig.~\ref{fig:bigpic}(c). Several LLVM passes are implemented to remove the strip-mining assemblies. Below is a detailed step-by-step explanation of the compilation process.

\textbf{Step 1:} The implementation of the built-in functions for Zoozve in Clang provides programmers with an intuitive interface to utilize Zoozve operations efficiently. These built-in functions allow developers to explicitly specify the vector value type, which is leveraged in subsequent optimizations. 

\textbf{Step 2:} Clang then transforms these built-in functions into intrinsics via intrinsic library mapping. Through static single assignment (SSA), variables in the high-level language are converted into virtual registers.

\textbf{Step 3:} To resolve the aforementioned compilation issue, the Zoozve intrinsic splitting pass is introduced. In this pass, the original intrinsic intermediate representation (IR) is transformed into a split form with a specific value type, ensuring that the range of virtual registers is clearly defined and effectively managed. The total number of split instances is determined by the split count. \texttt{delimiter} intrinsics are inserted before and after each split to guide the RA phase. These delimiters indicate which registers must be allocated consecutively to form a register group.

\textbf{Step 4:} Once the transformed IR is generated, it is utilized in the RA stage, where registers are assigned based on the lifespan of virtual registers as determined by live interval modification. The process operates between the lifetime analysis and RA, ensuring that virtual registers grouped by delimiter intrinsics are assigned consecutively. First, eligible registers are scanned, tracing back to locate the delimiter intrinsic and determine the $LMUL$. For each subsequent $LMUL$ split intrinsic, it enforces the lifespan of the registers to that of the first intrinsic. Finally, all split virtual registers are enqueued for RA in the designated queue. The RA operates by performing a lifetime analysis of virtual registers, allowing virtual registers grouped by delimiter intrinsics to be assigned in a manner that preserves their spatial continuity.

\textbf{Step 5:} Following RA, the $\text{IR}_{\text{split}}$ is translated into corresponding Zoozve assembly instructions (\texttt{Z\_asms}) through the Zoozve assembly coalescing pass. During this pass, consecutive \texttt{Z\_asms} are detected, where vector registers are consecutive and other parameters remain the same, and merged into a single, consolidated instruction, corresponding to the original built-in C function. 

\begin{figure}[!t]
    \centering
            \includegraphics[width=0.9\linewidth]{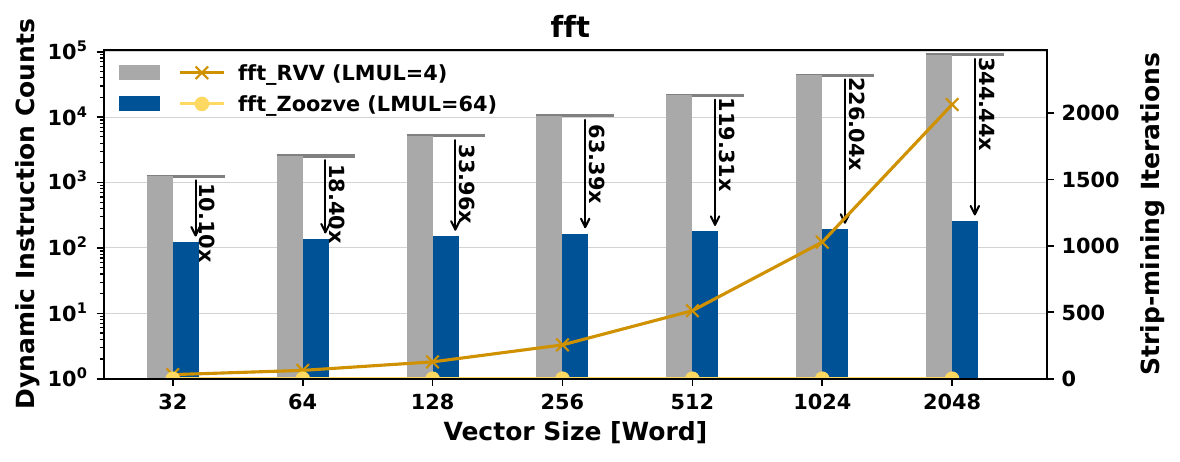}
            \includegraphics[width=0.9\linewidth]{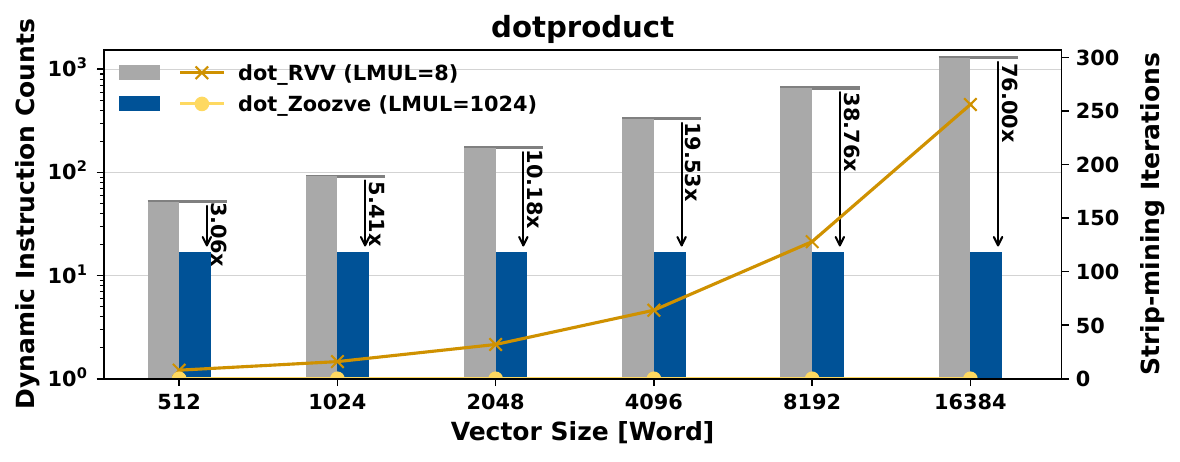}
            \includegraphics[width=0.9\linewidth]{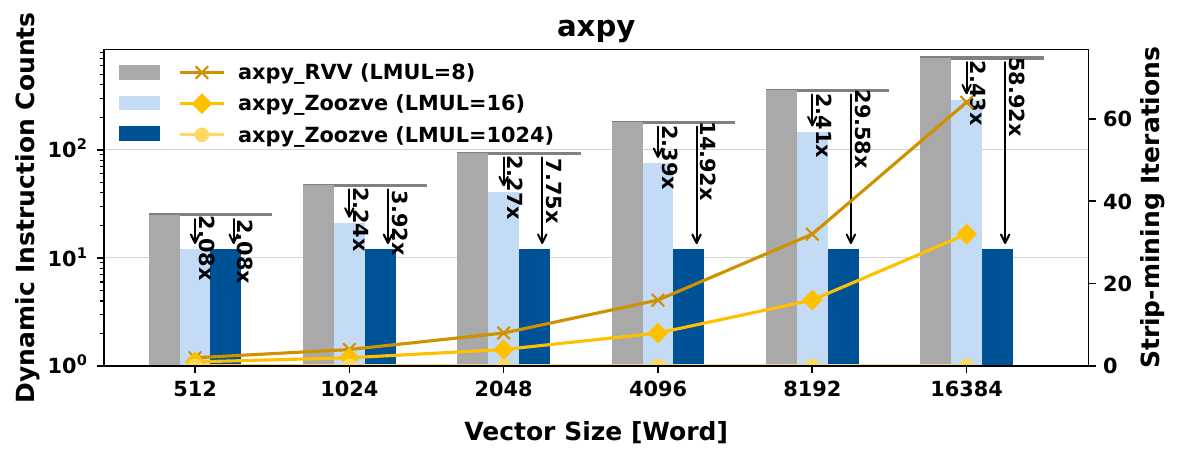}

    \vspace{-0.5cm}
    \caption{Speedup in dynamic instruction counts and strip-mining iterations for RVV and Zoozve across three kernels. }
    \label{fig:overall}
    \vspace{-0.7cm}
\end{figure}

\vspace{-0.3cm}
\section{Experiment and Evaluation}
\subsection{Experimental Setup}


We evaluate the performance of Zoozve against RVV using LLVM 15.6.0, the Spike simulator \cite{spike}, and a custom Zoozve simulator. Benchmarks include \texttt{dotproduct} and \texttt{axpy} from OpenBLAS \cite{openblas}, along with manually implemented \texttt{fft} kernels, all representative of scientific computing, signal processing, and machine learning workloads. The study examines dynamic instruction counts and scalable LMUL effects under configurations of up to 2048 registers and LMUL values of 1024. A hardware proof-of-concept is also implemented to demonstrate ISA feasibility.

\subsection{Kernel Comparison}

Fig.~\ref{fig:overall} compares the dynamic instruction count speedup and strip-mining iterations of Zoozve versus RVV across three kernels: \texttt{fft}, \texttt{dotproduct}, and \texttt{axpy}. In \texttt{fft}, Zoozve consistently reduces instruction counts and eliminates strip-mining overhead across data sizes from 32 to 2048 points. At 32 points, Zoozve achieves a 10.1$\times$ speedup, reaching 344.44$\times$ at 2048 points. Unlike the traditional divide-and-conquer approach \cite{cooley1965algorithm}, Zoozve avoids LMUL constraints, allowing efficient computation and permutation with minimal register spilling.

Zoozve also delivers substantial benefits in kernels like $\texttt{dotproduct}$ and $\texttt{axpy}$ ranging from 512 to 16,384. Both exhibit linear complexity $O(n)$, typically involving simple multiplication and addition operations. In RVV, limited register capacity causes instruction counts and strip-mining iterations to grow with data size -- for example, \texttt{dotproduct} sees instruction counts rise from 52 to 1292 and strip-mining from 8 to 256. Zoozve, with larger register files and flexible LMUL, maintains a constant instruction count of 17, eliminating strip-mining and achieving up to 76$\times$ speedup. A similar pattern occurs in \texttt{axpy}, where RVV’s instruction count increases from 25 to 707 and strip-mining count rises from 2 to 64, while Zoozve holds steady at 12 instructions, reaching a speedup of 58.92$\times$ at the largest data size.

\begin{figure}[!t]
    \centering
    \includegraphics[width=0.9\linewidth]{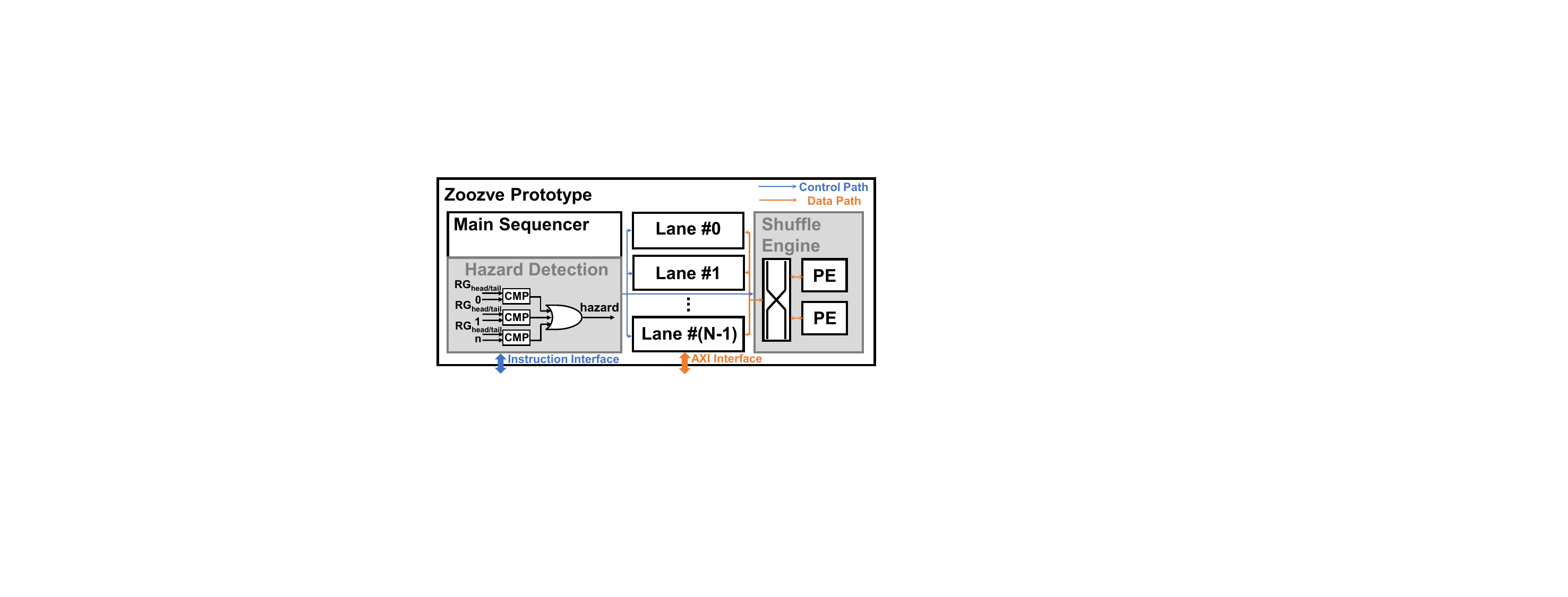}
    \vspace{-0.3cm}
    \caption{An architecture supporting the Zoozve extension, with the additional components required for Zoozve shaded.}
    \label{fig:arch}
    \vspace{-0.5cm}
\end{figure}

\begin{table*}[t]
  \centering
  \renewcommand\arraystretch{0.9}
  \caption{Files Generated by the Zoozve Compiler and Their Functions}
  \vspace{-0.4cm}
  \label{tab:artifact}
  \begin{threeparttable}[b]
  \begin{tabular}{ccccc}
    \toprule
    \# & Output file name & Executed by & Description & Correspond to \\
    \midrule
    \midrule
    1 & \texttt{venusbuiltin.h} & \texttt{python} & Generates a valid set of built-ins and formats. & Fig.~\ref{fig:bigpic}(c) Step 1 \\
    \midrule
    2 & \texttt{venustype.h} & \texttt{python} & Generates valid vector \texttt{typedef} declarations. & Fig.~\ref{fig:bigpic}(c) Step 1 \\
    \midrule
    3 & \texttt{test.0.ll} & \texttt{clang} & Generates normal IR from \texttt{test.c}. & Fig.~\ref{fig:bigpic}(c) Step 2 \\
    \midrule
    4 & \texttt{test.split.ll} & \texttt{opt} & Generates split IR from \texttt{test.0.ll}. & Fig.~\ref{fig:bigpic}(c) Step 3 \\ 
    \midrule
    \multirow{3}{*}{5} & \multirow{3}{*}{\texttt{test\_before\_merge.s}} & \multirow{3}{*}{\texttt{llc}} & Performs register allocation based on \texttt{test.split.ll} & \multirow{3}{*}{Fig.~\ref{fig:bigpic}(c) Step 4} \\ 
    &&&and generates the assembly file,& \\
    &&& but the registers are not coalesced.& \\
    \midrule
    \multirow{2}{*}{6} & \multirow{2}{*}{\texttt{test.s}} & \multirow{2}{*}{\texttt{llc}} & Performs register allocation based on \texttt{test.split.ll} & \multirow{2}{*}{Fig.~\ref{fig:bigpic}(c) Step 5} \\
    &&& and generates the assembly file with register coalescing.& \\
    \midrule
    7 & \texttt{test} & \texttt{clang} & Generates hardware executable file from \texttt{test.s}. & -- \\ 
    \midrule
    8 & \texttt{test\_asm.txt} & \texttt{llvm-objdump} & Generates disassembly file based on \texttt{test}. & -- \\
    \bottomrule
\end{tabular}
\end{threeparttable}
\end{table*}

\subsection{Hardware Proof-of-Concept}
Fig.~\ref{fig:arch} illustrates a possible hardware implementation for Zoozve. Building upon \cite{cavalcante2019ara}, two key components are introduced to support the 
 Zoozve extension. In the control path, additional logic is incorporated to accommodate the flexibility of the RG and detect hazards between instructions. Comparators (CMPs) determine whether register indices fall within the range of $RG_{head}$ and $RG_{tail}$, with their outputs OR'ed to generate a hazard signal. In the data path, a shuffle engine -- comprising a crossbar and multiple processing elements (PEs) -- is implemented to handle inter-lane asymmetric operations, while lanes execute symmetric operations. Our design is synthesized using the SMIC 40nm process (400 MHz), yielding a 7.2 mm$^2$ synthesis area and 11.9 mm$^2$ layout area for a 64-lane, 1024-register configuration, with a negligible 5.2\% area overhead.

\section{Conclusion}
This work presents Zoozve, a strip-mining-free RISC-V vector extension that tackles performance bottlenecks with arbitrary register grouping in ultra-long vector computation. Compiler optimizations, including intrinsic splitting and assembly coalescing, further enhance performance, achieving a 10.10$\times$ FFT instruction reduction with just a 5.2\% area increase. The source code and corresponding Docker environment are available at \cite{zoozvegithub} and \cite{xu_2025_15291421}.



\appendix

\section{Artifact Appendix}

\subsection{Abstract}

This appendix offers a comprehensive guide for interacting with the Zoozve compiler artifact. Our compiler is built on LLVM and incorporates a series of custom passes to enable arbitrary register grouping, allowing code to be compiled with significantly reduced instruction overhead. The following sections provide the environment setup process, after which users can compile and test the provided code, as well as reproduce and verify the results reported in the paper.


\subsection{Description \& Requirements}

\subsubsection{How to access}

Due to the large size of the artifact, we have provided three access methods for convenience:

\begin{itemize}
    \item GitHub Repository \cite{zoozvegithub}: This option allows users to compile the artifact on their own machine. The repository includes all necessary code and dependencies.

    \item Zenodo with Docker Environment \cite{xu_2025_15291421}: For a streamlined experience, we have also uploaded a pre-configured Docker image containing the entire compilation environment. This ensures that users can bypass manual setup and focus on the evaluation itself.

    \item \href{https://drive.google.com/drive/folders/1_ocKokjl7tiNqqzHFDmAJr-vl9Vnnm8v?usp=sharing}{Google Drive}: If neither of the above methods successfully reproduces the results, users can download the text files generated during the compilation process from Google Drive for evaluation.
\end{itemize}

\subsubsection{Hardware dependencies}
The Zoozve compiler does not impose any special hardware requirements and can be executed on a standard Linux system. Note that the current implementation runs only in debug mode and requires approximately 100 GB of available disk space.

\subsubsection{Software dependencies}
It is recommended to build and deploy the Zoozve compiler on Ubuntu 22.04. Our provided Docker environment is also based on this system.

\subsection{Installation}

The installation process can be found in the \href{https://github.com/ACELab-SHU/LCTES-25-Artifact/blob/main/README.pdf}{README} for reference on GitHub \cite{zoozvegithub}.

\subsection{Experiment Workflow}
\subsubsection{Major claims} 
Based on a specified set of \texttt{C} built-in functions, the Zoozve compiler can generate executable files for arbitrary-length \texttt{int8} and \texttt{int16} vector types. This workflow is illustrated in Fig.~\ref{fig:bigpic}(c) of the paper.

\subsubsection{Experiments} \textit{[Zoozve compiler evaluation][1 human-minute + 1 compute-minute + 10 MB disk]}
This experiment leverages the compiled executables of \texttt{clang}, \texttt{opt}, and \texttt{llc} to transform \texttt{C} source code into LLVM IR, and subsequently generate assembly and machine code. The complete workflow corresponds to Fig.~\ref{fig:bigpic}(c) in the paper.

{\setlength{\parindent}{0pt}
\textbf{How to:} To compile user code, run \texttt{make all} in the \texttt{zoozve/ venus\_test} directory.

\textbf{Results:} The compilation process generates both textual and binary output files, whose functions correspond to those described in the paper, as summarized in Table~\ref{tab:artifact}. For detailed information on artifact evaluation, please refer to the \href{https://github.com/ACELab-SHU/LCTES-25-Artifact/blob/main/README.pdf}{README} on GitHub \cite{zoozvegithub}.
}

\clearpage 
\bibliographystyle{ACM-Reference-Format}

\begin{thebibliography}{20}


\ifx \showCODEN    \undefined \def \showCODEN     #1{\unskip}     \fi
\ifx \showDOI      \undefined \def \showDOI       #1{#1}\fi
\ifx \showISBNx    \undefined \def \showISBNx     #1{\unskip}     \fi
\ifx \showISBNxiii \undefined \def \showISBNxiii  #1{\unskip}     \fi
\ifx \showISSN     \undefined \def \showISSN      #1{\unskip}     \fi
\ifx \showLCCN     \undefined \def \showLCCN      #1{\unskip}     \fi
\ifx \shownote     \undefined \def \shownote      #1{#1}          \fi
\ifx \showarticletitle \undefined \def \showarticletitle #1{#1}   \fi
\ifx \showURL      \undefined \def \showURL       {\relax}        \fi
\providecommand\bibfield[2]{#2}
\providecommand\bibinfo[2]{#2}
\providecommand\natexlab[1]{#1}
\providecommand\showeprint[2][]{arXiv:#2}

\bibitem[ACELab-SHU(2025)]%
        {zoozvegithub}
\bibfield{author}{\bibinfo{person}{ACELab-SHU}.} \bibinfo{year}{2025}\natexlab{}.
\newblock \bibinfo{title}{LCTES-25-Artifact}.
\newblock
\newblock
\newblock
\shownote{[Online]. Available: \url{https://github.com/ACELab-SHU/LCTES-25-Artifact}}.


\bibitem[Andri et~al\mbox{.}(2020)]%
        {andri2020extending}
\bibfield{author}{\bibinfo{person}{Renzo Andri}, \bibinfo{person}{Tomas Henriksson}, {and} \bibinfo{person}{Luca Benini}.} \bibinfo{year}{2020}\natexlab{}.
\newblock \showarticletitle{{Extending the RISC-V ISA for efficient RNN-based 5G radio resource management}}. In \bibinfo{booktitle}{\emph{57th ACM/IEEE Design Automation Conference (DAC)}}. \bibinfo{publisher}{IEEE Press}, \bibinfo{address}{Piscataway, NJ, USA}, \bibinfo{pages}{1--6}.
\newblock
\urldef\tempurl%
\url{https://doi.org/10.1109/DAC18072.2020.9218496}
\showDOI{\tempurl}


\bibitem[Cavalcante et~al\mbox{.}(2019)]%
        {cavalcante2019ara}
\bibfield{author}{\bibinfo{person}{Matheus Cavalcante}, \bibinfo{person}{Fabian Schuiki}, \bibinfo{person}{Florian Zaruba}, \bibinfo{person}{Michael Schaffner}, {and} \bibinfo{person}{Luca Benini}.} \bibinfo{year}{2019}\natexlab{}.
\newblock \showarticletitle{Ara: A {1-GHz+} scalable and energy-efficient {RISC-V} vector processor with multiprecision floating-point support in 22-nm {FD-SOI}}.
\newblock \bibinfo{journal}{\emph{IEEE Transactions on Very Large Scale Integration (VLSI) Systems}} \bibinfo{volume}{28}, \bibinfo{number}{2} (\bibinfo{year}{2019}), \bibinfo{pages}{530--543}.
\newblock
\urldef\tempurl%
\url{https://doi.org/10.1109/TVLSI.2019.2950087}
\showDOI{\tempurl}


\bibitem[Chen et~al\mbox{.}(2020)]%
        {chen2020xuantie}
\bibfield{author}{\bibinfo{person}{Chen Chen}, \bibinfo{person}{Xiaoyan Xiang}, \bibinfo{person}{Chang Liu}, \bibinfo{person}{Yunhai Shang}, \bibinfo{person}{Ren Guo}, \bibinfo{person}{Dongqi Liu}, \bibinfo{person}{Yimin Lu}, \bibinfo{person}{Ziyi Hao}, \bibinfo{person}{Jiahui Luo}, \bibinfo{person}{Zhijian Chen}, \bibinfo{person}{Chunqiang Li}, \bibinfo{person}{Yu Pu}, \bibinfo{person}{Jianyi Meng}, \bibinfo{person}{Xiaolang Yan}, \bibinfo{person}{Yuan Xie}, {and} \bibinfo{person}{Xiaoning Qi}.} \bibinfo{year}{2020}\natexlab{}.
\newblock \showarticletitle{Xuantie-910: A commercial multi-core 12-stage pipeline out-of-order 64-bit high performance {RISC-V} processor with vector extension: Industrial product}. In \bibinfo{booktitle}{\emph{ACM/IEEE 47th Annual International Symposium on Computer Architecture (ISCA)}}. \bibinfo{publisher}{IEEE Press}, \bibinfo{address}{Piscataway, NJ, USA}, \bibinfo{pages}{52--64}.
\newblock
\urldef\tempurl%
\url{https://doi.org/10.1109/ISCA45697.2020.00016}
\showDOI{\tempurl}


\bibitem[Cooley and Tukey(1965)]%
        {cooley1965algorithm}
\bibfield{author}{\bibinfo{person}{James~W Cooley} {and} \bibinfo{person}{John~W Tukey}.} \bibinfo{year}{1965}\natexlab{}.
\newblock \showarticletitle{An algorithm for the machine calculation of complex {Fourier} series}.
\newblock \bibinfo{journal}{\emph{Mathematics of computation}} \bibinfo{volume}{19}, \bibinfo{number}{90} (\bibinfo{year}{1965}), \bibinfo{pages}{297--301}.
\newblock
\urldef\tempurl%
\url{https://doi.org/10.2307/2003354}
\showDOI{\tempurl}


\bibitem[Gautschi et~al\mbox{.}(2017)]%
        {gautschi2017near}
\bibfield{author}{\bibinfo{person}{Michael Gautschi}, \bibinfo{person}{Pasquale~Davide Schiavone}, \bibinfo{person}{Andreas Traber}, \bibinfo{person}{Igor Loi}, \bibinfo{person}{Antonio Pullini}, \bibinfo{person}{Davide Rossi}, \bibinfo{person}{Eric Flamand}, \bibinfo{person}{Frank~K G{\"u}rkaynak}, {and} \bibinfo{person}{Luca Benini}.} \bibinfo{year}{2017}\natexlab{}.
\newblock \showarticletitle{Near-threshold {RISC-V} core with {DSP} extensions for scalable {IoT} endpoint devices}.
\newblock \bibinfo{journal}{\emph{IEEE transactions on very large scale integration (VLSI) systems}} \bibinfo{volume}{25}, \bibinfo{number}{10} (\bibinfo{year}{2017}), \bibinfo{pages}{2700--2713}.
\newblock
\urldef\tempurl%
\url{https://doi.org/10.1109/TVLSI.2017.2654506}
\showDOI{\tempurl}


\bibitem[Hennessy and Patterson(2011)]%
        {hennessy2011computer}
\bibfield{author}{\bibinfo{person}{John~L Hennessy} {and} \bibinfo{person}{David~A Patterson}.} \bibinfo{year}{2011}\natexlab{}.
\newblock \bibinfo{booktitle}{\emph{Computer Architecture, Fifth Edition: A Quantitative Approach}}.
\newblock \bibinfo{publisher}{Morgan Kaufmann Publishers Inc.}, \bibinfo{address}{340 Pine Street, Sixth Floor, San Francisco, CA, USA}.
\newblock
\urldef\tempurl%
\url{https://dl.acm.org/doi/10.5555/1999263}
\showURL{%
\tempurl}


\bibitem[Lai and Lee(2022)]%
        {lai2022efficient}
\bibfield{author}{\bibinfo{person}{Hung-Ming Lai} {and} \bibinfo{person}{Jenq-Kuen Lee}.} \bibinfo{year}{2022}\natexlab{}.
\newblock \showarticletitle{Efficient support of the scan vector model for {RISC-V} vector extension}. In \bibinfo{booktitle}{\emph{Workshop Proceedings of the 51st International Conference on Parallel Processing}}. \bibinfo{publisher}{Association for Computing Machinery}, \bibinfo{address}{New York, NY, USA}, \bibinfo{pages}{1--8}.
\newblock
\urldef\tempurl%
\url{https://doi.org/10.1145/3547276.3548518}
\showDOI{\tempurl}


\bibitem[Lomont(2011)]%
        {lomont2011introduction}
\bibfield{author}{\bibinfo{person}{Chris Lomont}.} \bibinfo{year}{2011}\natexlab{}.
\newblock \showarticletitle{Introduction to {Intel Advanced Vector Extensions}}.
\newblock \bibinfo{journal}{\emph{Intel white paper}}  \bibinfo{volume}{23} (\bibinfo{year}{2011}), \bibinfo{pages}{1--21}.
\newblock
\urldef\tempurl%
\url{https://hpc.llnl.gov/sites/default/files/intelAVXintro.pdf}
\showURL{%
\tempurl}


\bibitem[Louis et~al\mbox{.}(2019)]%
        {louis2019towards}
\bibfield{author}{\bibinfo{person}{Marcia~Sahaya Louis}, \bibinfo{person}{Zahra Azad}, \bibinfo{person}{Leila Delshadtehrani}, \bibinfo{person}{Suyog Gupta}, \bibinfo{person}{Pete Warden}, \bibinfo{person}{Vijay~Janapa Reddi}, {and} \bibinfo{person}{Ajay Joshi}.} \bibinfo{year}{2019}\natexlab{}.
\newblock \showarticletitle{Towards deep learning using {TensorFlow Lite} on {RISC-V}}. In \bibinfo{booktitle}{\emph{Third Workshop on Computer Architecture Research with RISC-V (CARRV)}}, Vol.~\bibinfo{volume}{1}. \bibinfo{pages}{6}.
\newblock
\urldef\tempurl%
\url{https://people.bu.edu/joshi/files/tflowlite-carrv-2019.pdf}
\showURL{%
\tempurl}

\newpage

\bibitem[Minervini et~al\mbox{.}(2023)]%
        {minervini2023vitruvius}
\bibfield{author}{\bibinfo{person}{Francesco Minervini}, \bibinfo{person}{Oscar Palomar}, \bibinfo{person}{Osman Unsal}, \bibinfo{person}{Enrico Reggiani}, \bibinfo{person}{Josue Quiroga}, \bibinfo{person}{Joan Marimon}, \bibinfo{person}{Carlos Rojas}, \bibinfo{person}{Roger Figueras}, \bibinfo{person}{Abraham Ruiz}, \bibinfo{person}{Alberto Gonzalez}, \bibinfo{person}{Jonnatan Mendoza}, \bibinfo{person}{Ivan Vargas}, \bibinfo{person}{C\'{e}sar Hernandez}, \bibinfo{person}{Joan Cabre}, \bibinfo{person}{Lina Khoirunisya}, \bibinfo{person}{Mustapha Bouhali}, \bibinfo{person}{Julian Pavon}, \bibinfo{person}{Francesc Moll}, \bibinfo{person}{Mauro Olivieri}, \bibinfo{person}{Mario Kovac}, \bibinfo{person}{Mate Kovac}, \bibinfo{person}{Leon Dragic}, \bibinfo{person}{Mateo Valero}, {and} \bibinfo{person}{Adrian Cristal}.} \bibinfo{year}{2023}\natexlab{}.
\newblock \showarticletitle{Vitruvius+: {An} area-efficient {RISC-V} decoupled vector coprocessor for high performance computing applications}.
\newblock \bibinfo{journal}{\emph{ACM Transactions on Architecture and Code Optimization}} \bibinfo{volume}{20}, \bibinfo{number}{2} (\bibinfo{year}{2023}), \bibinfo{pages}{1--25}.
\newblock
\urldef\tempurl%
\url{https://doi.org/10.1145/3575861}
\showDOI{\tempurl}


\bibitem[{NEC Corporation}(2018)]%
        {necaurora}
\bibfield{author}{\bibinfo{person}{{NEC Corporation}}.} \bibinfo{year}{2018}\natexlab{}.
\newblock \bibinfo{title}{{SX-Aurora TSUBASA} Architecture Guide Revision 1.1}.
\newblock
\newblock
\newblock
\shownote{[Online]. Available: \url{https://sxauroratsubasa.sakura.ne.jp/documents/guide/pdfs/Aurora_ISA_guide.pdf}}.


\bibitem[OpenMathLib({[n.\,d.]})]%
        {openblas}
\bibfield{author}{\bibinfo{person}{OpenMathLib}.} \bibinfo{year}{[n.\,d.]}\natexlab{}.
\newblock \bibinfo{title}{{OpenBLAS}}.
\newblock \bibinfo{howpublished}{GitHub. [Online]. Available: \url{https://github.com/OpenMathLib/OpenBLAS}}.
\newblock


\bibitem[Perotti et~al\mbox{.}(2024)]%
        {perotti2024ara2}
\bibfield{author}{\bibinfo{person}{Matteo Perotti}, \bibinfo{person}{Matheus Cavalcante}, \bibinfo{person}{Renzo Andri}, \bibinfo{person}{Lukas Cavigelli}, {and} \bibinfo{person}{Luca Benini}.} \bibinfo{year}{2024}\natexlab{}.
\newblock \showarticletitle{Ara2: Exploring Single-and Multi-Core Vector Processing with an Efficient {RVV} 1.0 Compliant Open-Source Processor}.
\newblock \bibinfo{journal}{\emph{IEEE Trans. Comput.}} \bibinfo{volume}{73}, \bibinfo{number}{7} (\bibinfo{year}{2024}), \bibinfo{pages}{1822--1836}.
\newblock
\urldef\tempurl%
\url{https://doi.org/10.1109/TC.2024.3388896}
\showDOI{\tempurl}


\bibitem[Perotti et~al\mbox{.}(2022)]%
        {perotti2022new}
\bibfield{author}{\bibinfo{person}{Matteo Perotti}, \bibinfo{person}{Matheus Cavalcante}, \bibinfo{person}{Nils Wistoff}, \bibinfo{person}{Renzo Andri}, \bibinfo{person}{Lukas Cavigelli}, {and} \bibinfo{person}{Luca Benini}.} \bibinfo{year}{2022}\natexlab{}.
\newblock \showarticletitle{A `{New Ara}' for vector computing: An open source highly efficient {RISC-V V} 1.0 vector processor design}. In \bibinfo{booktitle}{\emph{IEEE 33rd International Conference on Application-specific Systems, Architectures and Processors (ASAP)}}. \bibinfo{publisher}{IEEE Press}, \bibinfo{address}{Piscataway, NJ, USA}, \bibinfo{pages}{43--51}.
\newblock
\urldef\tempurl%
\url{https://doi.org/10.1109/ASAP54787.2022.00017}
\showDOI{\tempurl}


\bibitem[Pohl et~al\mbox{.}(2019)]%
        {pohl2019performance}
\bibfield{author}{\bibinfo{person}{Angela Pohl}, \bibinfo{person}{Mirko Greese}, \bibinfo{person}{Biagio Cosenza}, {and} \bibinfo{person}{Ben Juurlink}.} \bibinfo{year}{2019}\natexlab{}.
\newblock \showarticletitle{A performance analysis of vector length agnostic code}. In \bibinfo{booktitle}{\emph{International Conference on High Performance Computing \& Simulation (HPCS)}}. \bibinfo{publisher}{IEEE Press}, \bibinfo{address}{Piscataway, NJ, USA}, \bibinfo{pages}{159--164}.
\newblock
\urldef\tempurl%
\url{https://doi.org/10.1109/HPCS48598.2019.9188238}
\showDOI{\tempurl}


\bibitem[Schuiki et~al\mbox{.}(2020)]%
        {schuiki2020stream}
\bibfield{author}{\bibinfo{person}{Fabian Schuiki}, \bibinfo{person}{Florian Zaruba}, \bibinfo{person}{Torsten Hoefler}, {and} \bibinfo{person}{Luca Benini}.} \bibinfo{year}{2020}\natexlab{}.
\newblock \showarticletitle{Stream semantic registers: A lightweight {RISC-V ISA} extension achieving full compute utilization in single-issue cores}.
\newblock \bibinfo{journal}{\emph{IEEE Trans. Comput.}} \bibinfo{volume}{70}, \bibinfo{number}{2} (\bibinfo{year}{2020}), \bibinfo{pages}{212--227}.
\newblock
\urldef\tempurl%
\url{https://doi.org/10.1109/TC.2020.2987314}
\showDOI{\tempurl}


\bibitem[Spike({[n.\,d.]})]%
        {spike}
\bibfield{author}{\bibinfo{person}{Spike}.} \bibinfo{year}{[n.\,d.]}\natexlab{}.
\newblock \bibinfo{title}{Spike: {RISC-V ISA} Simulator}.
\newblock \bibinfo{howpublished}{GitHub. [Online]. Available: \url{https://github.com/riscv/riscv-isa-sim}}.
\newblock


\bibitem[Stephens et~al\mbox{.}(2017)]%
        {stephens2017arm}
\bibfield{author}{\bibinfo{person}{Nigel Stephens}, \bibinfo{person}{Stuart Biles}, \bibinfo{person}{Matthias Boettcher}, \bibinfo{person}{Jacob Eapen}, \bibinfo{person}{Mbou Eyole}, \bibinfo{person}{Giacomo Gabrielli}, \bibinfo{person}{Matt Horsnell}, \bibinfo{person}{Grigorios Magklis}, \bibinfo{person}{Alejandro Martinez}, \bibinfo{person}{Nathanael Premillieu}, \bibinfo{person}{Alastair Reid}, \bibinfo{person}{Alejandro Rico}, {and} \bibinfo{person}{Paul Walker}.} \bibinfo{year}{2017}\natexlab{}.
\newblock \showarticletitle{The {ARM Scalable Vector Extension}}.
\newblock \bibinfo{journal}{\emph{IEEE micro}} \bibinfo{volume}{37}, \bibinfo{number}{2} (\bibinfo{year}{2017}), \bibinfo{pages}{26--39}.
\newblock
\urldef\tempurl%
\url{https://doi.org/10.1109/MM.2017.35}
\showDOI{\tempurl}


\bibitem[Xu et~al\mbox{.}(2025)]%
        {xu_2025_15291421}
\bibfield{author}{\bibinfo{person}{Siyi Xu}, \bibinfo{person}{Limin Jiang}, \bibinfo{person}{Yintao Liu}, \bibinfo{person}{Yihao Shen}, \bibinfo{person}{Yi Shi}, \bibinfo{person}{Shan Cao}, {and} \bibinfo{person}{Zhiyuan Jiang}.} \bibinfo{year}{2025}\natexlab{}.
\newblock \bibinfo{booktitle}{\emph{(Artifact) Zoozve: A Strip-Mining-Free RISC-V Vector Extension with Arbitrary Register Grouping Compilation Support (WIP)}}.
\newblock
\urldef\tempurl%
\url{https://doi.org/10.5281/zenodo.15291421}
\showDOI{\tempurl}


\end{thebibliography}

\end{document}